\documentclass[epj]{svjour_2}
\usepackage{graphicx}
\usepackage{appendix}

\usepackage{amsmath}
\usepackage{amsfonts}
\usepackage{amssymb}

\begin{document}

\title{Multiplicity Studies and Effective Energy in ALICE at the LHC}



\author{
A. Akindinov\inst{1},
A.~Alici\inst{2,3},
P.~Antonioli\inst{3},
S.~Arcelli\inst{2,3},
M.~Basile\inst{2,3},
G.~Cara~Romeo\inst{3},
M.~Chumakov\inst{1},
L.~Cifarelli\inst{2,3},
F.~Cindolo\inst{3},
A.~De~Caro\inst{5},
D.~De~Gruttola{\inst5},
S.~De~Pasquale\inst{5},
M.~Fusco~Girard\inst{5},
C.~Guarnaccia\inst{5},
D.~Hatzifotiadou\inst{3},
H.T.~Jung\inst{6},
W.W.~Jung\inst{6},
D.W.~Kim\inst{6},
H.N.~Kim\inst{6},
J.S.~Kim\inst{6,7},
S.~Kiselev\inst{1},
G.~Laurenti\inst{3},
K.~Lee\inst{6},
S.C.~Lee\inst{6},
E.~Lioublev\inst{1},
M.L.~Luvisetto\inst{3},
A.~Margotti\inst{3},
A.~Martemiyanov\inst{1},
R.~Nania\inst{3},
F.~Noferini\inst{2,3,4},
P.~Pagano\inst{5},
A.~Pesci\inst{3},
R.~Preghenella\inst{2,3},
G.~Russo\inst{5},
E.~Scapparone\inst{3},
G.~Scioli\inst{2,3},
R.~Silvestri\inst{5},
Y.~Sun\inst{7},
I.~Vetlitskiy\inst{1},
K.~Voloshin\inst{1},
L.~Vorobiev\inst{1},
M.C.S.~Williams\inst{3},
B.~Zagreev\inst{1},
C.~Zampolli\inst{2,3,4},
A.~Zichichi\inst{2,3,4}}
\institute{
Institute for Theoretical and Experimental Physics, Moscow, Russia \and
Dipartimento di Fisica dell'Universit\`a, Bologna, Italy \and 
Sezione INFN, Bologna, Italy 
\and Museo Storico della Fisica e Centro Studi e Ricerche ``Enrico
Fermi'', Rome, Italy
\and Dipartimento di Fisica dell'Universit\`a and INFN, Salerno, Italy \and
Department of Physics, Kangnung National University, Kangnung, 
South Korea \and World Laboratory, Lausanne, Switzerland}

\vspace{1cm}

\dedication{
In memory of A.~Smirnitskiy\inst{1}
}

\abstract{
In this work we explore the possibility to perform ``effective energy'' 
studies in very high energy collisions at the CERN Large Hadron Collider (LHC). In particular,
we focus on the possibility to measure in $pp$ collisions the average charged multiplicity as a function of the effective energy with the ALICE experiment, using its capability to measure the energy of the leading baryons with the Zero Degree Calorimeters. Analyses of this kind have been done at lower centre--of--mass energies and have shown that, once the appropriate kinematic variables are chosen, particle production 
is characterized by universal properties: no matter the nature of the interacting particles, the final states have identical features.
Assuming that this universality picture can be extended to {\it ion--ion} collisions, as suggested by recent results from RHIC experiments,
a novel approach based on the scaling hypothesis for limiting fragmentation has been used to derive the expected charged event 
multiplicity in $AA$ interactions at LHC. 
This leads to scenarios where 
the multiplicity is significantly lower compared to most of the predictions from the models currently used to describe high energy $AA$ collisions.
A mean charged multiplicity of about $1000-2000$ per rapidity unit (at $\eta \sim 0$) is expected for the most central $Pb-Pb$ collisions at $\sqrt{s_{NN}} = 5.5~{\rm TeV}$.
\PACS{
      {PACS-key}{discribing text of that key}   \and
      {PACS-key}{discribing text of that key}
     } 
}
\maketitle
\section{Introduction}
\label{sec:Intro}
In high--energy particle collisions, bulk event properties 
like the average charged particle multiplicity are
regarded as experimental observables of fundamental interest, 
providing important information on the dynamics of the interaction.
In particular, the average charged particle multiplicity in multihadronic 
final states has so far been measured in many different 
interaction systems ($e^+e^-$ and $pp(\bar{p})$
collisions, DIS processes, etc.) and over a wide range 
of centre--of--mass energies. Although the data show a dependence on 
$\sqrt{s}$  which is characteristic of the 
specific initial state under consideration,
as pointed out in~\cite{Basile:ManyPaper,ISRdata_ref,Basile:1982we} 
a universal behaviour can actually be identified if the appropriate 
definition of the energy available 
for particle production (the ``effective energy'') is used. 

The aim of this work is to address the possibility to perform 
an effective energy study at energies of several TeV at LHC, 
with the ALICE experiment~\cite{ALICE,ALICEaddendum1,ALICEaddendum2}. 
With this respect ALICE (A Large Ion Collider Experiment) 
has an excellent capability,
thanks to the presence of several detectors for the measurement 
of the particle multiplicity over a wide rapidity range~\cite{Carminati:2004fp,Carminati:PPR2}. Moreover,
on both sides of the beam interaction point, the detector will be 
equipped with very forward calorimeters, the Zero Degree Calorimeters (ZDCs) \cite{ZDC:1999ke}, which will allow to derive the 
effective energy on an event--by--event basis by measuring 
the energy of the leading nucleons.

This paper is organized as follows. 
In section~\ref{sec:ExpData} we briefly review the main experimental 
results which support the existence of a universal behaviour in 
particle production, independently of the nature of the colliding system. 
In section~\ref{sec:PythiaData} a simplified analysis 
based on the PYTHIA Monte Carlo is presented, showing that this 
event generator is able to reproduce to a good extent 
the experimental observations described in section~\ref{sec:ExpData},
and can therefore be used to evaluate the feasibility of 
an effective energy study at the LHC energies. 
A more detailed analysis is then described in section~\ref{ALICEdet}, 
where the capability of the ALICE detector to 
measure the energy of the leading baryons in $pp$ collisions at the LHC 
is quantified using a realistic simulation of the ALICE ZDC detector response. 
Finally, in section~\ref{Nucleus_nucleus} we present 
a prediction for the total charged multiplicity in $Pb-Pb$ interactions 
at LHC, assuming that the universality features 
discussed herein hold also for ultrarelativistic heavy ion collisions.

\section{What we learned from previous experiments}
\label{sec:ExpData}

\begin{figure}[!h]
  \small
  \begin{center}
    \hspace{-0.4cm}
    \includegraphics[width=1.1\linewidth]{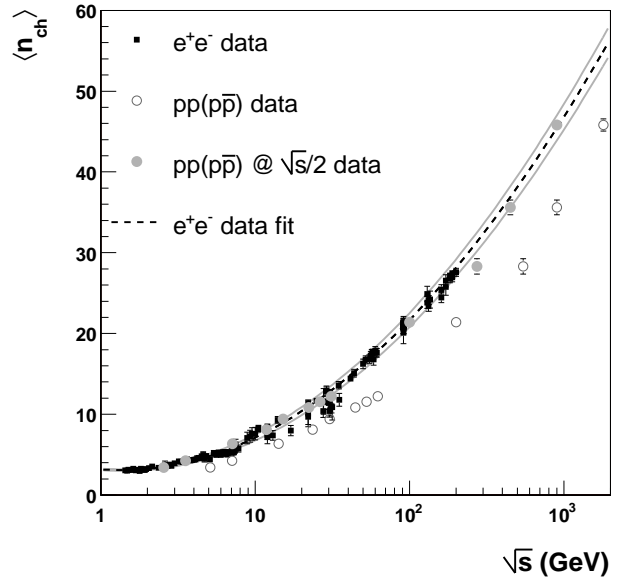}
    \caption{Average charged particle multiplicity measured in $e^+e^-$  
    (full black symbols) and $pp(\bar{p})$ 
 (open symbols) collisions. A logarithmic fit to 
  the $e^+e^-$ data is also shown as the dashed line. The full grey symbols indicate the
$pp(\bar{p})$ data after the centre--of--mass energies have been scaled by 
a factor $\frac{1}{2}$ (see text).}
    \label{fig:eeppcomb}
  \end{center}
\end{figure}

It is well known that the average charged multiplicity 
 in $e^+e^-$ collisions follows a logarithmic dependence on 
the centre--of--mass energy $\sqrt{s}$. 
Figure~\ref{fig:eeppcomb} shows a compilation of data from $e^+e^-$ 
experiments \cite{Niczyporuk:1981kf,Alam:1982ue,Bartel:1983qp,Abrams:1989rz,Derrick:1986jx,Aihara:1986mv,Zheng:1990iq,aleph,Abreu:1997ir,Adeva:1991it,opal,net_ee} over a wide range of centre--of--mass energies (full black symbols), together  
with the result of a logarithmic fit to the measurements, indicated by the dashed line.
As mentioned in section~\ref{sec:Intro}, the average charged multiplicity 
is characterized by a significantly different 
dependence on $\sqrt{s}$ if other initial states  
are considered. In particular, in the case of $pp(\bar{p})$ 
collisions the charged particle multiplicity 
at a fixed centre--of--mass energy is observed to be systematically 
lower than what can be inferred from the $e^+e^-$ data at 
the same $\sqrt{s}$ (see again Fig.~\ref{fig:eeppcomb}, open symbols, 
data from~\cite{pp_ref,Biyajima:2001ud,UA5:1984-87,E735:1998-99,Abe:1993xy}). 
As pointed out in \cite{Basile:ManyPaper}, this behaviour can be 
understood after considering that, while in $e^+e^-$ collisions the 
energy available for particle production coincides with the full centre--of--mass energy
(once the effects from the initial state radiation are removed), 
in $pp(\bar{p})$ collisions this energy
is reduced with respect to $\sqrt{s}$ 
due to a basic feature of the hadronic interactions, the ``leading effect''.

The leading effect, which is related 
to the quantum number flow between the initial and the final state,
implies that in  $pp(\bar{p})$ collisions there is a high 
probability to emit a forward baryon with large longitudinal momentum
along the direction of the incident beams. Since the leading baryon 
carries away a fraction of the incident energy, the energy 
available for particle production is reduced with respect to the total
centre--of--mass energy. 
If the leading effect is taken into account in the definition of the 
effective energy available for particle production, $E_{eff}$, common 
universal features can be observed in the two different interaction 
systems, $e^+e^-$ and $pp(\bar{p})$. 
 
In particular, as done in \cite{Basile:1982we}, the effective energy 
can be estimated on an event--by--event basis by measuring the energy $E_{leading}$ 
of the leading baryon in each event hemisphere (the hemispheres being defined
with respect to a plane transverse to the direction of the incident beams in the centre--of--mass system). The effective energy 
per hemisphere $(E_{eff})_{i}$ is then given by:

\begin{equation}
(E_{eff})_{i}~=~\sqrt{s}/2~-~(E_{leading})_{i} \hspace{1cm}\textrm{for } i=1,2.
\end{equation}

Notice that, by measuring the energy of the leading particles event 
by event, a wide range of effective energies can be covered at a 
fixed centre--of--mass energy. 
Relying on the experimental observation of the independence of the 
two event hemispheres with respect to the leading effect \cite{Basile:1982we}, 
the total effective energy $E_{eff}$ available for particle production in the 
whole event can be derived from the measurement of only one leading particle per event, using the relation: 
\begin{eqnarray}
\label{eq:Eeff1}
 E_{eff} = 2 (E_{eff})_{i}  =  \sqrt{s} - 2 (E_{leading})_{i}.
\end{eqnarray}
Correspondingly, the average total charged multiplicity in the event is given by :
\begin{equation}
  \left<n_{ch}\right> = 2 \left<n_{ch}\right>_{i}.
\end{equation}
 Alternatively, in those events where a leading particle is measured in 
both hemispheres, the total effective energy can also be derived from the relation:
\begin{equation}
\label{eq:Eeff2}
  E_{eff}  = \sqrt{s~[1~-~(x_{F})_{1}][1~-~(x_{F})_{2}]},
\end{equation}
where $(x_{F})_{i}$ are the measured Feynman--x of the 
two leading nucleons, $(x_{F})_{i}~=~2~(p_L)_{i}/ \sqrt{s}$.

The two methods of defining the effective energy in the event 
have been proved to be equivalent,
as shown in~\cite{Basile:ManyPaper}; it should be noticed, however, 
that the first one has the advantage of making optimal use of the 
data in terms of the collected event statistics, obviously owing to the much larger acceptance and efficiency of single with respect to double leading particle detection.

\begin{figure}[!h]
  \small
  \begin{center}
    \includegraphics[width=\linewidth]{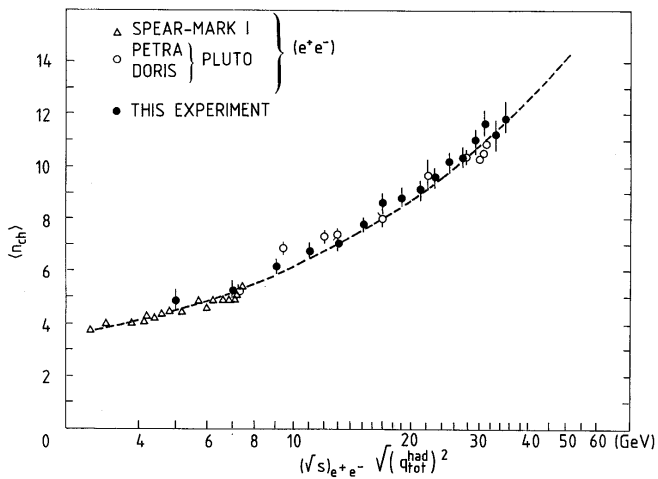}
    \caption{ Average charged multiplicity $\left<n_{ch}\right>$ 
     as a function of the effective 
      energy $E_{eff}$ (here indicated as $\sqrt{(q^{had}_{tot})^2}$),
      as measured in minimum bias $pp$ collisions collected 
      by the SFM experiment at the CERN ISR (full circles). 
      The data from $e^+ e^-$ experiments are also shown 
      (open circles and triangles) in terms of $\sqrt{s}$. A fit to
     ISR--$pp$ data is superimposed. The plot is taken from~\cite{ISRdata_ref}.}
    \label{pp_efflead}
  \end{center}
\end{figure}

As mentioned before, after taking into account the leading effect 
and using the effective energy instead of the full centre--of--mass energy, a very good agreement between the measured charged particle multiplicity in $e^+e^-$ and $pp$ collisions is obtained, as shown in Fig.~\ref{pp_efflead}.
The $pp$ data in this figure refer to a ``minimum bias'' event sample from which elastic and diffractive processes were excluded, with leading protons in the range $0.3 \lesssim x_{F} \lesssim 0.8$ (see later on sections \ref{sec:PythiaData} and \ref{ALICEdet}).

A good agreement between the inclusive $pp(\bar{p})$ results presented 
in Fig.~\ref{fig:eeppcomb} and $e^+e^-$ results is also obtained 
if the $pp(\bar{p})$ centre--of--mass energies are 
scaled by a factor $\frac{1}{2}$ (full grey symbols). 
This is a direct consequence of the effective energy being the relevant 
variable in particle production, and of the assumption, supported by 
several measurements \cite{Basile:ManyPaper}, that 
the Feynman--x distribution of the leading particles has a mean value equal to $\sim 0.5$, resulting in an average effective energy which is about half the full centre--of--mass energy.

\section{Pythia simulations}
\label{sec:PythiaData}
The PYTHIA6.214~\cite{Sjostrand:2006za} Monte Carlo has been 
used to simulate $pp$ collisions at a centre--of--mass energy of $14$~TeV.
An effective energy study carried on for $pp$ collisions requires a precise definition of the type of events to be used. Following the choice of the original ISR--$pp$ experiment \cite{Basile:ManyPaper,ISRdata_ref,Basile:1982we}, only non diffractive minimum bias events have been considered herein.
It should be noted that most of the other $pp(\bar{p})$ experiments \cite{pp_ref,Biyajima:2001ud,UA5:1984-87,E735:1998-99,Abe:1993xy} quoted in section~\ref{sec:ExpData} included double--diffractive processes in their definition of minimum bias.
With PYTHIA the double--diffraction contribution is at the level of $15\%$ with respect to the non diffractive one at $\sqrt{s} = 14~{\rm TeV}$ (using default generation parameters).
A brief description of the ALICE $pp$ minimum bias trigger capabilities for both diffractive and non diffractive processes is given in subsection~\ref{ALICEtrigger}.

To optimize the agreement with respect to the dependence observed 
in the data for the charged particle multiplicity as a function of the effective energy, some of the PYTHIA parameters regulating the treatment of the multiple parton interactions (MI) have been tuned.
This MI model has been introduced and widely discussed in the literature \cite{Nakada:99lhcb,Field:APP,Moraes:2006tune,Acosta:CMS06} in order to describe the basic features and, in particular, the charged particle multiplicity of high energy minimum bias $pp(\bar{p})$ final states up to $\sqrt{s} \sim 1~{\rm TeV}$. It is still under investigation but represents however one possible tool to perform extrapolations in the multi--{\rm TeV} domain at LHC.
The set of values used in the simulation is reported in table~\ref{tab:pythia}.

\begin{table}[h]
\begin{center}
\begin{tabular}{|c|c|c|}
\hline
Parameter & Our tuning & Default\\
\hline
MSTP(51) & 7 (CTEQ5L) & 7 (CTEQ5L)\\
MSTP(81) & 1    & 1\\
MSTP(82) & 2    & 4\\
PARP(82) & 2.15 GeV & 1.80 GeV\\  
PARP(89) & 1000 GeV & 1000 GeV\\ 
PARP(90) & 0.215 & 0.160\\
\hline
\end{tabular}
\end{center}
\caption{Values of the PYTHIA parameters tuned for this work to generate minimum bias (non diffractive) $pp$ events at various centre--of--mass energies.}
\label{tab:pythia}
\end{table} 

Notice that the only three parameters tuned correspond, respectively, to the selection of the MI mechanism features in terms of hadronic matter distribution (MSTP(82)), the regularization scale of the transverse momentum spectrum of the process (PARP(82)) which controls the MI rate and the exponent of the $\sqrt{s}$--power law dependence of the MI mechanism (PARP(90)) \cite{Sjostrand:2006za}.

\begin{figure}[!h]
  \small
  \begin{center}
    \includegraphics[width=1.1\linewidth]{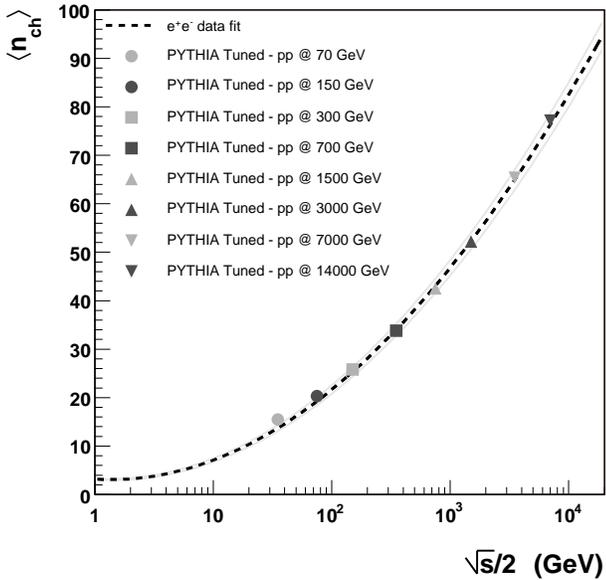}
    \caption{Mean charged multiplicity $\left<n_{ch}\right>$ in $pp$ collisions from PYTHIA Monte Carlo, at $\sqrt{s}~=~70$ to $14000$~GeV. The energy on the horizontal axis is scaled by a factor $\frac{1}{2}$ in order to take into account an average leading effect in the $pp$ final states. The $e^+e^-$ fit of Fig.~\ref{fig:eeppcomb} is also plotted.}
    \label{fig:pythiaTuned}
  \end{center}
\end{figure}

Figure~\ref{fig:pythiaTuned} shows the mean total charged 
multiplicity for generated $pp$ collisions over a wide range 
of centre--of--mass energies, with $\sqrt{s}$ rescaled by a factor 
$\frac{1}{2}$ to take into account, on average,  
the energy carried away by the leading nucleons, 
as done in section~\ref{sec:ExpData}. 
To derive the average charged multiplicity $\left<n_{ch}\right>$ 
all primary charged particles in the event
(including the decay products of short--lived resonances) with no transverse momentum cutoff
and in the full angular acceptance have been considered.
Although $\left<n_{ch}\right>$ is derived from the Monte Carlo information at the event generator level, notice that the ALICE detector is expected to measure the multiplicity with a good accuracy, high efficiency and over a wide acceptance both in case of $pp$ and $Pb-Pb$ collisions (as reported  in chapter 6.1.5 of the ALICE Physics Performance Report vol. II  \cite{Carminati:PPR2}). 

Using the tuned PYTHIA Monte Carlo, a very good agreement with the fit that describes the $\sqrt{s}$ dependence 
of the multiplicity for $e^{+}e^{-}$ collisions 
(see Fig.~\ref{fig:eeppcomb}) is observed in Fig.~\ref{fig:pythiaTuned} over the whole range of energies.
Using instead the default values, PYTHIA disagrees by $\sim 10\%$ with the $e^{+}e^{-}$ fit for $\sqrt{s} > 1~{\rm TeV}$. 

\begin{figure}[!h]
  \small
  \begin{center}
    \includegraphics[width=1.1\linewidth]{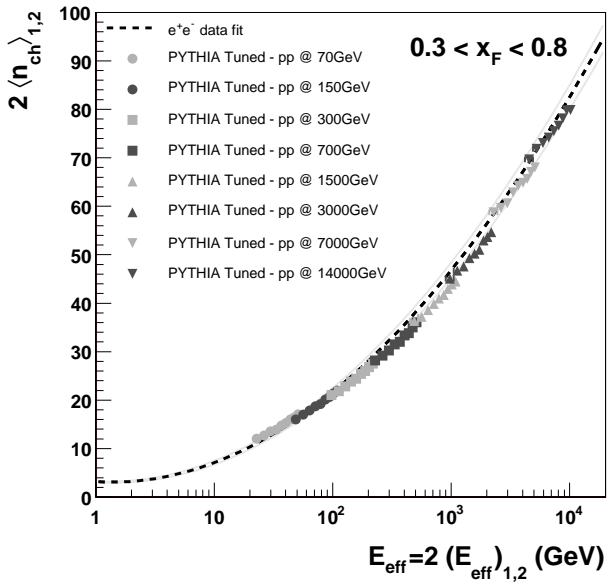}
    \caption{Mean charged multiplicty $\left<n_{ch}\right>$ in
      $pp$ PYTHIA events at $\sqrt{s} = 70$ to $14000$~GeV with respect to effective energy. The $e^{+}e^{-}$ fit is also plotted. Here the single hemisphere variables are considered and scaled by a factor of 2 (see text).}
    \label{pythiaTunedEffectiveEnergy}
  \end{center}
\end{figure}

A satisfactory agreement is also observed when studying the multiplicity 
as a function of the effective energy in each single event. Following 
the approach described in section~\ref{sec:Intro}, for each event the effective energy in a single hemisphere was derived from the energy of the corresponding leading baryon (defined as the highest longitudinal momentum baryon in the hemisphere) and then multiplied by 2, according to relation (\ref{eq:Eeff1}). 
The total charged multiplicity was also derived on a single hemisphere basis, scaled by the same factor of 2.
As can be seen from Fig.~\ref{pythiaTunedEffectiveEnergy}, also in 
this case the dependence of the total charged multiplicity on the effective 
energy follows closely\footnote{The values deviate systematically from the $e^{+}e^{-}$ data fit as $E_{eff}$ increases whereas they always start well on the fit line. This effect is likely due to the fact that the multiple interaction parametrization of PYTHIA generator slightly depends also on the centre-of-mass energy.} the fit to the $e^{+}e^{-}$ data. 

\begin{figure}[!h]
  \small
  \begin{center}
    \includegraphics[width=1.1\linewidth]{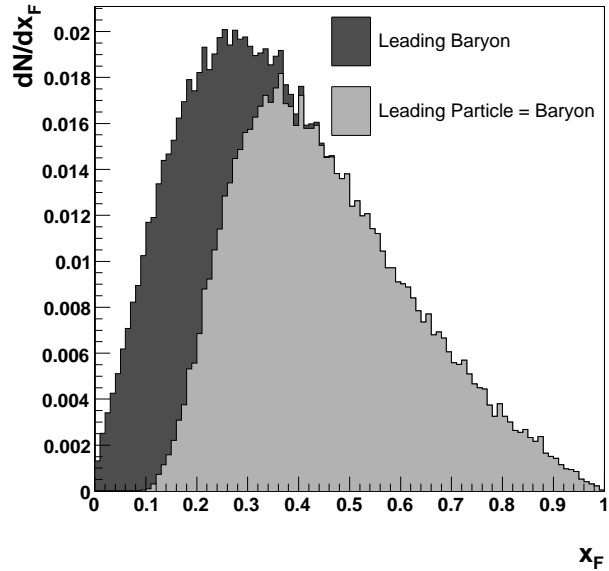}
    \caption{ The $x_{F}$ distribution of the most energetic baryon (dark--shaded histogram) in each hemisphere for $pp$ collisions at $\sqrt{s} = 14$~TeV. The $x_{F}$ distribution 
of this baryon in the case it is also the most energetic particle in the 
hemisphere is shown as the light--shaded histogram.}
    \label{pythiaLeadingNucleon_1}
  \end{center}
\end{figure}

Notice that in the above analysis the Feynman--x of the leading baryons was restricted to be in the range:
$0.3<x_{F}<0.8.$
The lower cutoff, $x_{F}^{min}=0.3$, is 
mainly motivated by the fact that in a realistic analysis 
the identity of the leading particle will not be accessible.
Some care is therefore required in defining an appropriate 
Feynman--x acceptance, in order 
to ensure that the energy measured in the forward region is indeed 
in most cases due to a leading baryon, and that the contamination 
from leading particles which are not baryons is kept below a reasonably 
small level. 
In Fig.~\ref{pythiaLeadingNucleon_1} the $x_{F}$ distribution 
of the leading baryon, defined as the proton/neutron with the highest 
longitudinal momentum (i.e. with the highest Feynman--x  
$x_{F} = 2 p_L/ \sqrt{s}$) among all baryons produced in a given event hemisphere, is shown as the 
dark--shaded histogram, for 10$^{5}$  $pp$ collisions generated at $\sqrt{s}=14$~TeV.
The $x_{F}$ distribution for the cases where this leading nucleon 
is also the leading particle among all particles (baryons or non baryons) of the same hemisphere is superimposed as the light--shaded histogram. 
For Feynman--x greater than 0.3, it can be seen that the leading nucleon 
is indeed the leading particle in the event hemisphere.
Moreover, a request of a minimum $x_{F}$ allows to reduce 
the contamination from leading particles that are not baryons.
At $x_{F} = 0.3$ this contamination is $\sim 30\%$ and rapidly decreases for $x_{F} > 0.3$.

For what concerns the constraint on the maximum $x_{F}$, the
requirement $x_{F}^{max} = 0.8$ was ispired by the original effective energy data analysis \cite{Basile:ManyPaper,ISRdata_ref,Basile:1982we}, where the same kind of cutoff on leading protons was imposed. This $x_{F} < 0.8$ condition, which fairly matches the acceptance of the ALICE ZDC calorimeters for leading baryons (see later on section~\ref{ALICEdet}), would also imply the rejection of the proton\footnote{For neutrons single--diffraction processes with charge exchange are highly unlike.} single--diffraction peak (not considered in this study).

\begin{figure}[!h]
  \small
  \begin{center}
    \includegraphics[width=1.1\linewidth]{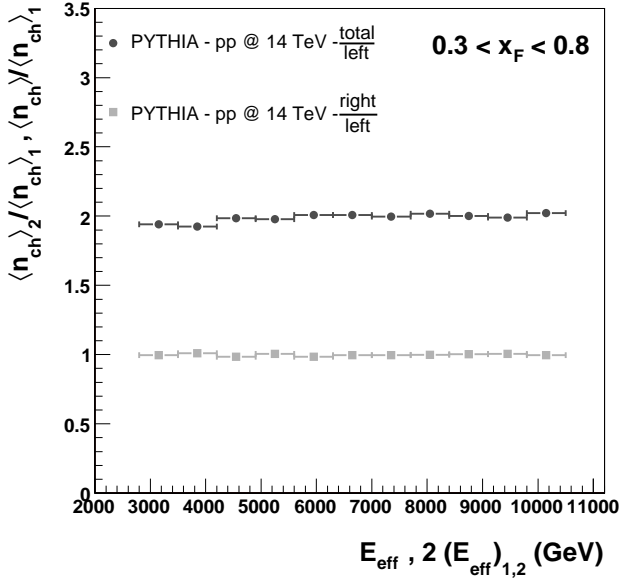}
    \caption{ Average charged multiplicity ratio $\left<n_{ch}\right> / \left<n_{ch}\right>_{1}$  as a function of the total effective energy $E_{eff}$ (black symbols) and average charged multiplicity ratio $\left<n_{ch}\right>_{2} / \left<n_{ch}\right>_{1}$
as a function of the effective energy  in a single hemisphere 
scaled by a factor $2$, $E_{eff}=2 (E_{eff})_{1,2}$ (grey symbols).}
    \label{pythiaHemisphereMultiplicity}
  \end{center}
\end{figure}

The PYTHIA generator reproduces to a good extent also other features 
related to the leading effect which have been observed in the data. As an 
example, Fig.~\ref{pythiaHemisphereMultiplicity} shows, for $pp$ 
collisions at $\sqrt{s}=14$~TeV, the ratio between 
the total charged multiplicity  $\left<n_{ch}\right>$, measured as 
a function of the total effective energy $E_{eff}$ given by equation (\ref{eq:Eeff2}), and the average 
charged multiplicity in a single hemisphere, measured as a function of 
the effective energy in a single hemisphere, 
scaled by a factor $2$, $E_{eff}=2 (E_{eff})_{1,2}$. 
The ratio is consistent with the relation  $\left<n_{ch}\right> = 2 \left<n_{ch}\right>_{1,2}$
in agreement with the hypothesis of the two event 
hemispheres being independent with respect to the leading effect,
and of the equivalence of the two 
effective energy $E_{eff}$ definitions
presented in section~\ref{sec:Intro}, equations~(\ref{eq:Eeff1}) 
and ~(\ref{eq:Eeff2}). 

\section{Tagging of leading nucleons with the ALICE ZDC detector in $pp$ collisions}
\label{ALICEdet}

Although the programme of the ALICE experiment has its main focus 
on heavy ion physics, the detector will also have 
excellent physics capabilities in the case of $pp$ collisions.
In particular, this applies to the analysis we are presenting
in this paper.

\subsection{ALICE performance for the selection and measurement of $pp$ events}
\label{ALICEtrigger}
A detailed description of the ALICE trigger system can be found elsewhere \cite{Carminati:PPR2,triggerALICE}. Here we will limit ourselves to a brief report on both the minimum bias trigger selection and the charged particle multiplicity measurement \cite{Carminati:PPR2} that ALICE will be able to perform in $pp$ collisions at LHC.

The minimum bias trigger is realized coupling different signals coming from the V0 \cite{V0ref} and the SPD \cite{SPDref} detectors.
The general purpose of this minimum bias trigger is to select events from $pp$ collisions with an efficiency as high as possible, and a bias as low as possible, and to reject events due to beam--gas or beam--halo.
The V0 is made up of two different arrays of scintillators placed along the beam pipe in the forward/backward directions.
It provides two kinds of signals: (i) VZERO.OR that requires at least one hit in one counter on one side and (ii) VZERO.AND that requires at least one hit in one counter on both sides.
The SPD (Silicon Pixel Detector) is made up of two coaxial layers of silicon sensors which also provide a trigger signal (GLOB.FO) consisting in a fast--OR signal of all its 1200 chips (400 in the inner layer, 800 in the external one).
Different combinations of these signals, coupled to a non background trigger (notBG, defined by requiring no signals in either of the V0 counters within appropriate time windows corresponding to beam--background processes), set different trigger configurations.
The three configurations are listed in table~\ref{tab:TriggerList}, while the corresponding efficiencies are reported in table~\ref{tab:Trigger}. 

\begin{table}[h]
\begin{tabular}{|l|l c c c r|}
\hline
MB 1 & GLOB.FO & or & VZERO.OR & and & notBG\\
MB 2 & GLOB.FO & and & VZERO.OR & and & notBG\\
MB 3 & GLOB.FO & and & VZERO.AND& and & notBG\\
\hline
\end{tabular} 
\begin{tabular}{c}
\\
\end{tabular}
\caption{Signal configurations for $pp$ minimum bias trigger.}
\label{tab:TriggerList}
\end{table} 

\begin{table}[h]
\begin{tabular}{|l|c|c|c|}
\hline
Process ($\sigma_{14{\rm TeV}}~{\rm [mb]}$) & MB1 (\%) & MB2 (\%) & MB3 (\%)\\
\hline
Non Diff. (55.22) & 99.9 & 99.1 & 96.9\\
Single--Diff. (14.30) & 73.8 & 59.5 & 38.4\\
Double--Diff. (9.78) & 87.8 & 68.7 & 45.6\\
\hline
\end{tabular} 
\begin{tabular}{c}
\\
\end{tabular}
\caption{Minimum bias trigger efficiency for different (diffractive and non diffractive) $pp$ processes.}
\label{tab:Trigger}
\end{table} 

The multiplicity can be measured counting the ``tracklets'' (extracted by the association of the signals in both the two layers of the SPD) with the SPD in the central region and with the FMD (Forward Multiplicity Detector) \cite{V0ref} in the forward region.
The pseudorapidity range covered by the ALICE detector is about 8 $\eta$--units.
In this range an accurancy on the multiplicity better than $10\%$ with $0.1~\eta$--units bin width can be achieved 
(as estimated in single central HIJING \cite{Wang:1991ht} events).

In the following we will assume that the uncertainty on the multiplicity measurement is negligible with respect to that on the effective energy measurement.

\subsection{ZDC performance for leading nucleons}
\label{ZDCperform}
As mentioned in section~\ref{sec:Intro}, 
the ALICE experiment will be able to measure the energy of the leading 
particles needed for this study thanks to the presence of the Zero Degree Calorimeters (ZDCs) \cite{ZDC:1999ke}. The aim of this section is to show that this system, although designed for the detection 
of nearly monochromatic ($E_{N}\approx 2.75$~TeV) 
leading nucleons in ion--ion collisions at $\sqrt{s_{NN}}=5.5$~TeV,  
will be able to measure on an event--by--event 
basis leading particles in $pp$ collisions over a wide range of energies
and with a good accuracy. 

\begin{figure}[!h]
  \small
  \begin{center}
    \includegraphics[width=\linewidth]{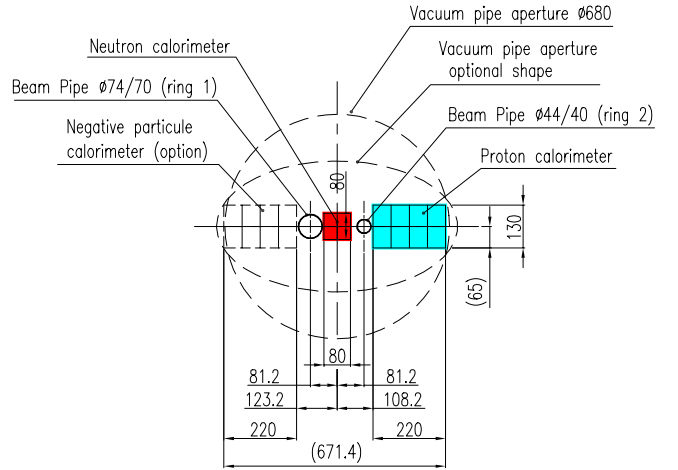}
    \caption{Transverse section of the LHC beam line at a distance of 
    $115~{\rm m}$ from the interaction vertex. The 
     location of the two ZDC calorimeters for protons and neutrons is also 
     shown.}\label{zdcSection}
  \end{center}
\end{figure}

A scheme of the ZDC apparatus is shown in Fig.~\ref{zdcSection}; it consists of two identical elements, placed on both sides of 
the interaction vertex at a distance of $115~{\rm m}$. 
Each element consists of two distinct calorimeters, one for 
leading neutrons (ZDCN), placed at
zero degrees relative to the LHC axis, and one for leading protons (ZDCP), 
placed externally to the outgoing beam pipe
on the side where positively charged particles are deflected by the 
beam optics. For charged particles, the Feynman--x range 
experimentally accessible with the ZDCs
will be $x_{F}\in [0.30-0.64]$, as the beam optics selects energies
between $2.2$~TeV and $4.5$~TeV. For neutral particles such 
constraints are clearly not present.
A more detailed description of the calorimeters is given in~\cite{ZDC:1999ke}.

The performance of the ALICE ZDC calorimeters were first checked 
on single highly energetic protons and neutrons,
with generated longitudinal momenta in the range 
between $1$~TeV and $6$~TeV.
As mentioned before, due to the beam optics
the ZDC acceptance is restricted to protons with energies in 
the range  $2.2<E<4.5$~TeV, while 
in the case of neutrons the whole range of energy is accessible.
The GEANT \cite{Brun:1987ma} package and  
the ALICE simulation and reconstruction software, AliRoot~\cite{AliRoot},
were used for the simulation of the detector response, and to perform realistic
digitization and reconstruction in the ZDC calorimeters.

\begin{figure}[!h]
  \small
  \begin{center}
    \includegraphics[width=\linewidth]{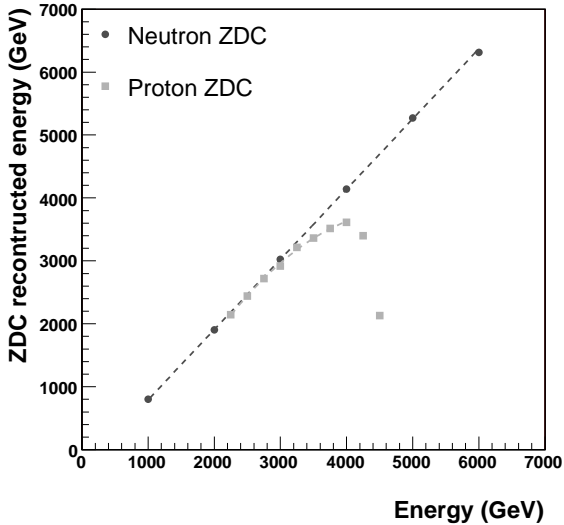}
    \caption{Reconstructed energy in the proton (grey symbols) and neutron (black symbols) ZDC calorimeters, as a function of the proton/neutron generated energy.}
    \label{fig:zdcEnergy}
  \end{center}
\end{figure}

\begin{figure}[!h]
  \small
  \begin{center}
    \includegraphics[width=\linewidth]{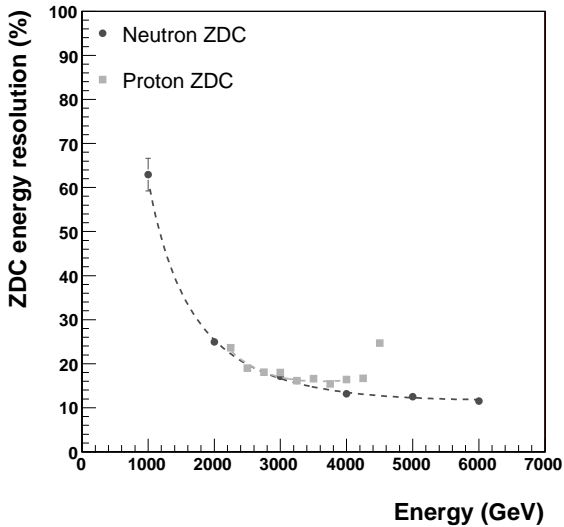}
    \caption{Energy resolution $\delta E / E$ of the proton (grey symbols)
    and neutron (black symbols) ZDC calorimeters, as a function of the 
     proton (neutron) generated energy.} 

    \label{fig:zdcResolution}
  \end{center}
\end{figure}

In Fig.~\ref{fig:zdcEnergy} the energy reconstructed in the ZDCs
for protons and neutrons as a function of the generated energy is shown.
While a good linearity is observed in the case of neutrons over the whole range of energies, in the case of protons the reconstructed energy 
is underestimated with respect to the true energy above $\sim 3$~TeV. 
This is because with increasing energy, the impact point of 
these particles falls closer
and closer to the edge of the detector due to the beam optics,
and the hadron cascade is no longer fully contained in the calorimeter.
The energy resolution, shown in 
Fig.~\ref{fig:zdcResolution}, ranges between $10\%$ and $20 \%$ 
both in the neutron and proton cases in the selected energy range (from $2.2$ to $4.5~{\rm TeV}$).
The absolute error on the energy measurement is about $500$~GeV.
For protons the ZDCP energy resolution could likely be improved using the information of the impact point of the track on the calorimeter front face.

\begin{figure}[!h]
  \small
  \begin{center}
   \includegraphics[width=6cm]{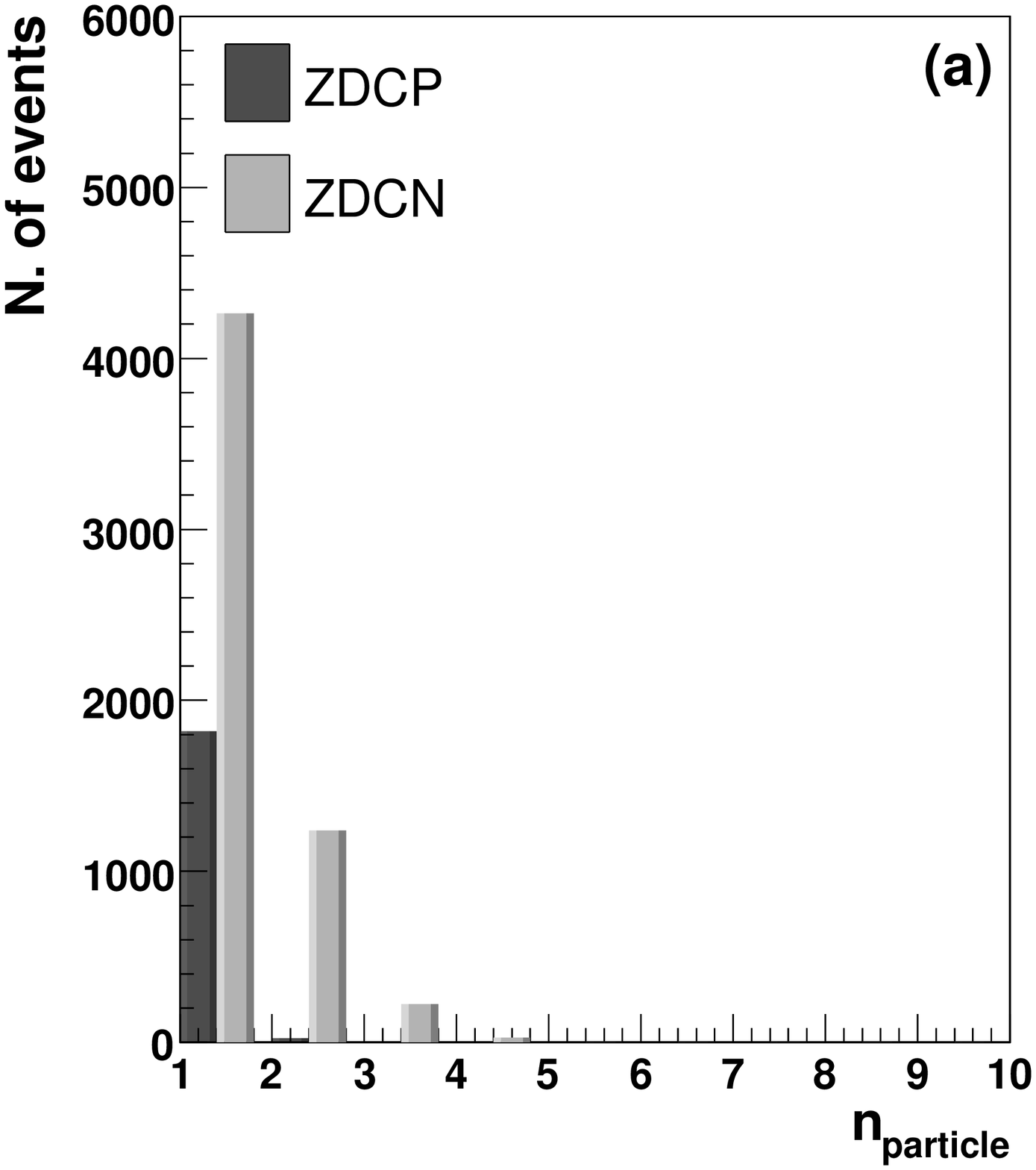} 
   \includegraphics[width=6cm]{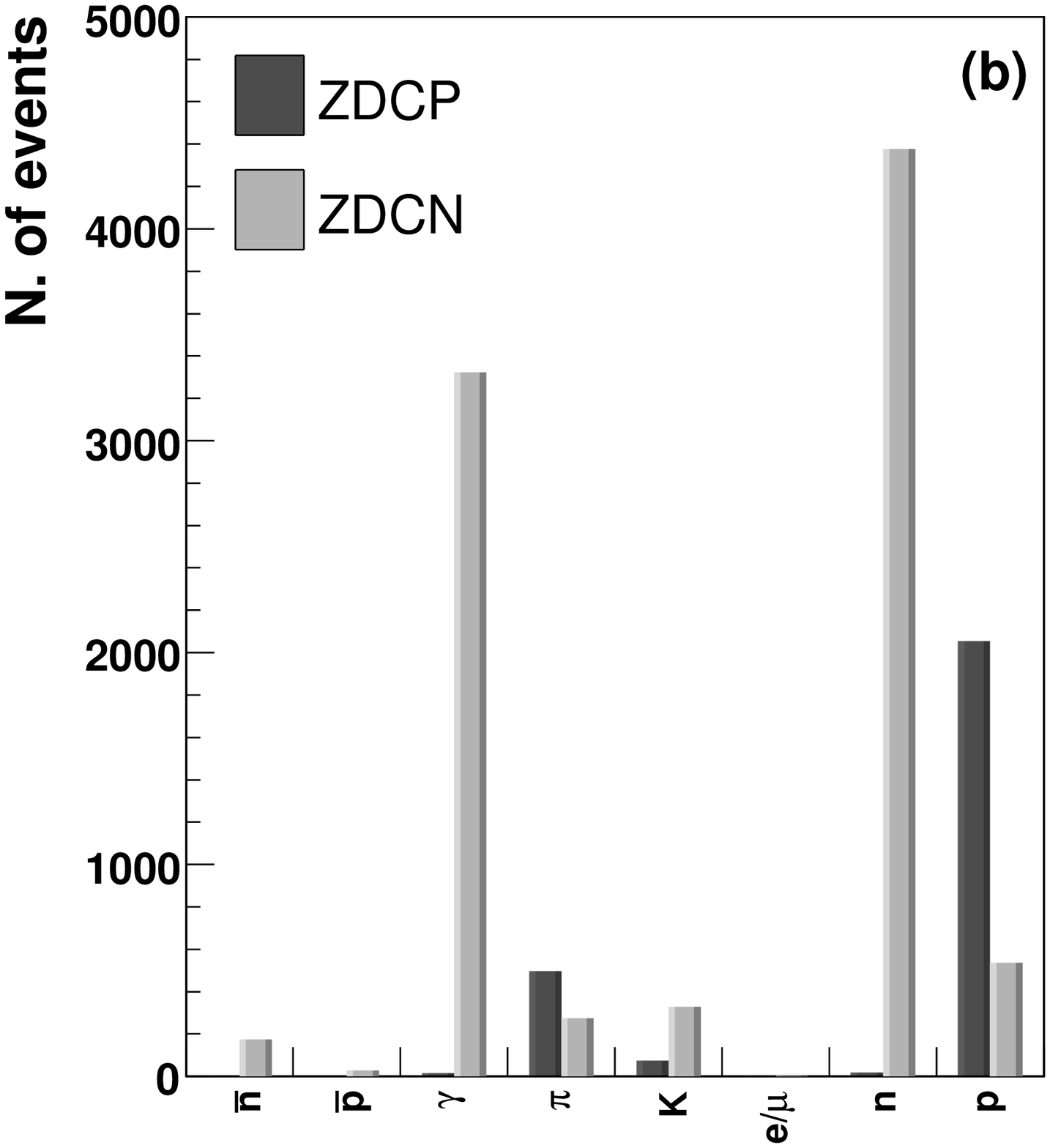}
   \caption{Multiplicity of particles ($n_{particle}$) detected in each section of the ZDC calorimeters (a) and the corresponding particle species composition (b), per single hemisphere.}
    \label{fig:ZDCMult}
  \end{center}
\end{figure}

\begin{figure}[!h]
  \small
  \begin{center}
    \includegraphics[width=\linewidth]{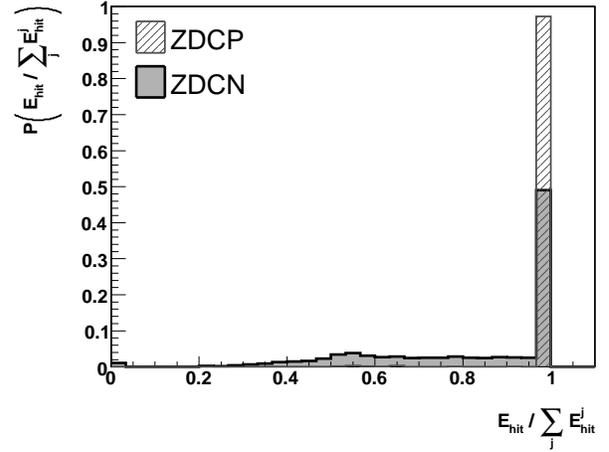}
    \caption{Probability distribution with respect to the fraction of deposited energy of the fastest particle in the ZDCP or ZDCN.}
    \label{fig:ZDCfrac}
  \end{center}
\end{figure}

After checking that the ZDCs can measure the energy of 
forward--going nucleons over an extended $x_{F}$ range, 
their performance has been tested on fully simulated $pp$ collisions.
Starting from 14000 generated events, in about $75\%$ of the events 
a signal is recorded in either the proton (ZDCP) or 
the neutron (ZDCN) calorimeter, in the forward or backward regions. 
In the majority of cases the proton calorimeter is hit by a 
single particle, while in the case 
of the neutron calorimeter the multiplicity is higher, due to 
additional photons coming from the decay of high--energy forward 
$\pi^{0}$ mesons produced in the interaction (in the case of 
the proton calorimeter, the corresponding contamination
which would be expected from charged pions 
is to a large extent removed 
by the beam optics). This is demonstrated
in Fig.s~\ref{fig:ZDCMult}.a) and ~\ref{fig:ZDCMult}.b),
where the multiplicity of particles 
detected in each of the ZDC calorimeters\footnote{In order to reject very low--energy background, only particles 
with Feynman--x $x_F>0.1$ were considered.} 
and their particle species composition are respectively shown. 
Moreover, looking at the fraction of deposited energy  
of the most energetic particle 
hitting the ZDCs (Fig~\ref{fig:ZDCfrac}), it can be seen
that in nearly all cases 
more than $95\%$ of the total energy released in the ZDCP is due 
to a single particle, while in the ZDCN this happens only in 
about $50\%$ of the cases. This is again mostly due to the background 
coming from photons from $\pi^{0}$ decays mentioned above.
 
\begin{figure}[!h]
  \begin{center}
    \includegraphics[width=\linewidth]{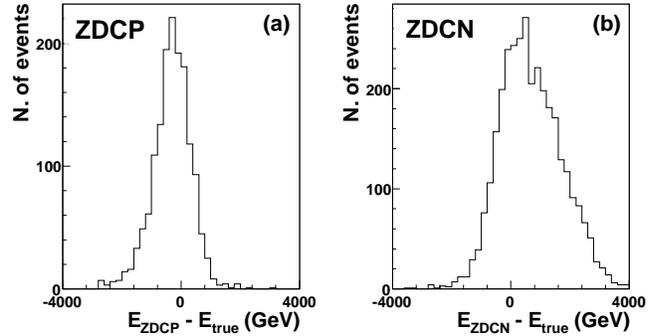}
    \caption{Difference between the reconstructed and the generated energy in the ZDCP (a) and ZDCN (b). In the case of the ZDCN, the clear distortion on the positive side of the distribution in (b) is due to the presence of high energy $\gamma$'s from $\pi^0$ decays,
 which cause an overestimate of the reconstructed energy.}
    \label{fig:ZDCEffEnergyTotalRes}
  \end{center}
\end{figure}

The presence of additional particles impinging on the neutron calorimeter 
causes a substantial distorsion when the total energy measured 
in the calorimeter is used to estimate the energy of the leading particle
entering the ZDCN, as can be seen in Fig~\ref{fig:ZDCEffEnergyTotalRes}.b,
where the difference between the total energy measured in the ZDCN 
calorimeter and the generated leading particle energy is shown. 

In the case of the proton calorimeter, where this type of background 
is much lower, the distribution is nearly gaussian 
(see Fig~\ref{fig:ZDCEffEnergyTotalRes}.a), although the 
energy is, on average, slightly underestimated due to hadronic showers 
not being fully contained for $E>3$~TeV.
Therefore, in terms of the effective energy reconstruction, 
the proton calorimeter ZDCP is expected to provide a cleaner measurement 
with respect to the ZDCN. On the other hand, 
the neutron calorimeter covers a range in 
Feynman--x which is more extended than in the case of the proton calorimeter, 
as discussed before in this section.

\begin{figure}[!h]
  \small
  \begin{center}
    \includegraphics[width=\linewidth]{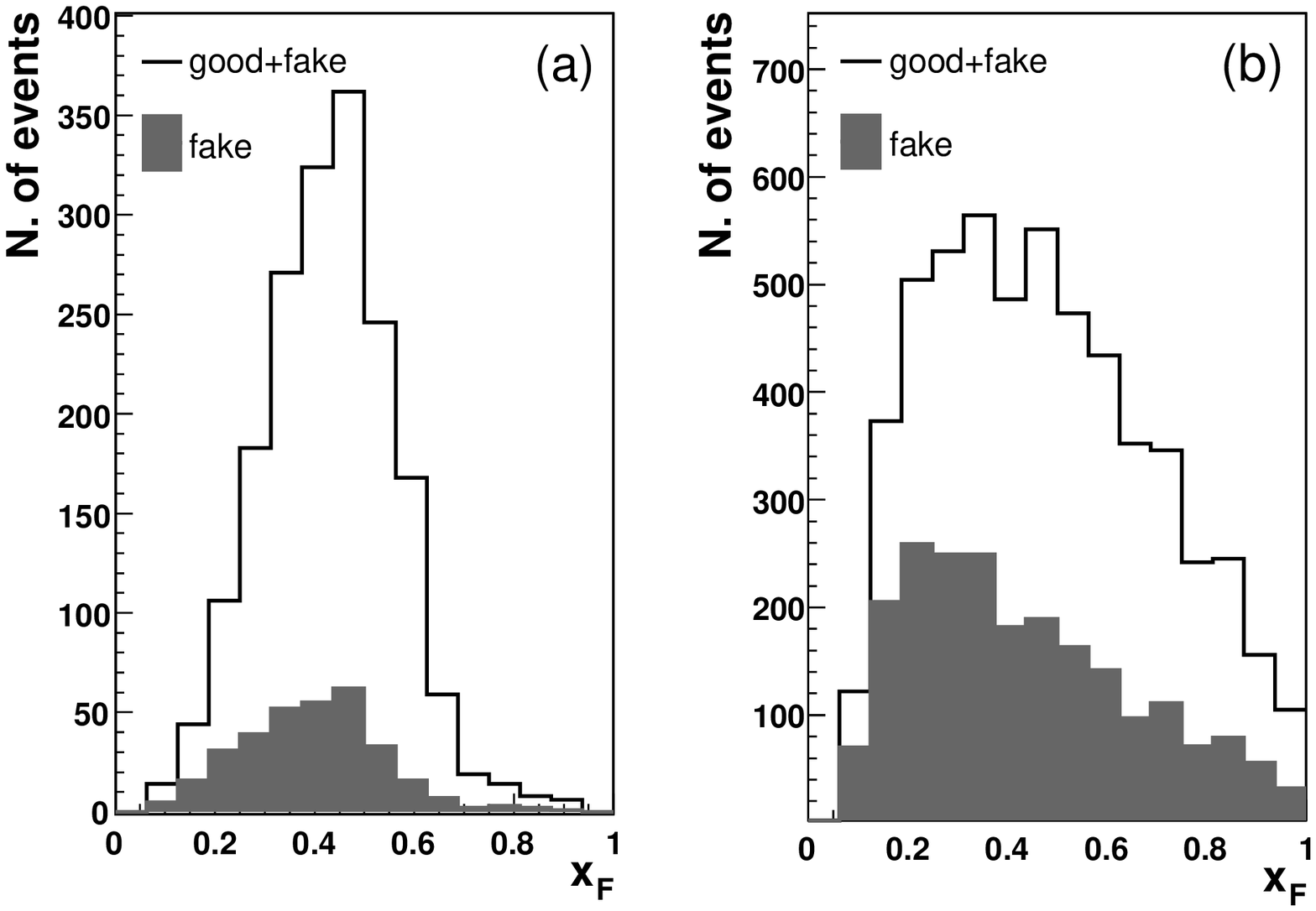}
    \caption{Distribution of the leading Feynman--x reconstructed 
with the ZDCP (a) and ZDCN (b) (empty histograms). The 
contamination from the cases where the highest energy deposit in 
the calorimeter was not due to a leading baryon are shown as dark shaded 
histograms.}
    \label{fig:ZDCxF}
  \end{center}
\end{figure}

The reconstructed $ x_{F}^{reco}=2 E_{ZDC}/\sqrt{s}$ distribution, 
where $E_{ZDC}$ is the total energy in either the ZDCP or
the ZDCN, is shown in Fig~\ref{fig:ZDCxF}. 

\begin{figure}[!h]
  \small
  \begin{center}
    \includegraphics[width=1.1\linewidth]{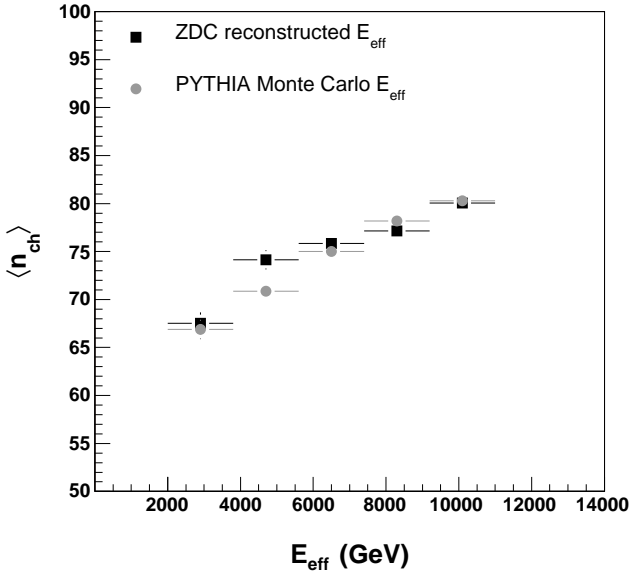}
    \caption{Average charged multiplicity as a function of the effective energy reconstructed in the ZDC calorimeters (black symbols) and of the effective energy calculated at the generator level (grey symbols).}
    \label{zdcFinalResults}
  \end{center}
\end{figure}

In those cases where a signal is detected in both ca\-lo\-rimeters, 
the highest among the two energies was taken as an estimate of the 
leading particle energy. The empty and the 
dark--shaded histograms indicate respectively the Feynman--x distribution of the reconstructed leading particles and the contamination from 
cases where the highest energy deposit in the calorimeter was not due to a leading baryon. This latter component, which can be regarded as a
source of  contamination in the measurement of the effective energy, 
is about 30\% in the case of the ZDCP and 40\% in the case of the ZDCN
over the Feynman--x range used in the analysis ($0.3< x_{F}^{reco}<0.8$).
This contamination, however, is expected not to affect significantly the 
quality of the measurement of the multiplicity as a fuction 
of the effective energy.
This is shown in Fig.~\ref{zdcFinalResults}, 
where the generated average charged event multiplicity 
as a fuction of the reconstructed
effective energy  $E_{eff}^{reco} = \sqrt{s} - 2  E_{ZDC}$ (black symbols), and of the effective energy calculated at the generator level (grey symbols) are compared. In both cases equation (\ref{eq:Eeff1}) was used.
The width of the horizontal binning is determined by the energy resolution of the ZDC calorimeters. 

The results are consistent, indicating that 
the limitations related to the reconstruction of the effective 
energy in the ZDCs and the background from other particles in the 
forward region do not introduce a significant bias in the measurement.
We therefore expect that the ALICE experiment will have a good capability 
to perform an effective energy study in $pp$ collisions at LHC.

\section{Nucleus--nucleus collisions}
\label{Nucleus_nucleus}
As described in the previous sections, hadron--hadron collisions produce a mean charged multiplicity in good agreement with the $e^+e^-$ collisions at the same effective energy.
Another question is still open. What happens in nucleus--nucleus?
The aim of this section is to investigate the possible scenario 
which we expect to be observed at LHC.

\begin{figure}[!h]
  \small
  \begin{center}
    \includegraphics[width=\linewidth]{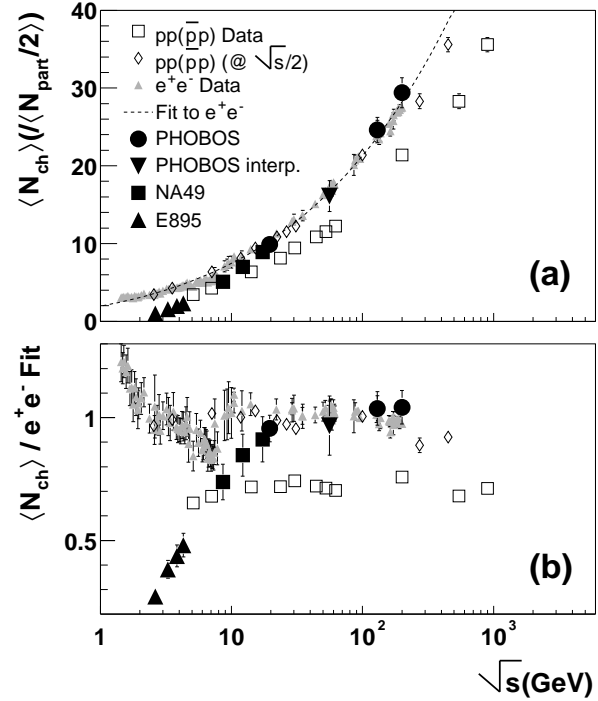}
    \caption{(a) Average charged multiplicity for $pp$, $p\bar{p}$, $e^+e^-$  and central $Au-Au$ events as a function of the centre--of--mass energy per binary collision. The $Au-Au$ ($Pb-Pb$ for NA49 experiment) data are scaled by a factor $\left< N_{part}/2 \right>$. The dotted line is a perturbative QCD fit to the $e^+e^-$ data. The diamonds are the $pp(\bar{p})$ data plotted at energy values scaled by a factor $1/2$. (b) The data in (a) divided by the $e^+e^-$ fit, to allow direct comparison of different data at the same energy. Both plots (a) and (b) are taken from \cite{AA:mult}.}   
    \label{AAenergyScale}
  \end{center}
\end{figure}

The PHOBOS experiment at RHIC has recently shown (Fig.~\ref{AAenergyScale}) how the total charged multiplicity in $AA$ collisions, scaled by the number of participant nucleons, varies with $\sqrt{s_{NN}}$ energy \cite{AA:mult}.
In particular, the $AA$ data indicate that the scaled total charged multiplicity does not depend, to a good approximation, on the centrality of the collision and that 
its value vs. $\sqrt{s_{NN}}$, for $20 < \sqrt{s_{NN}} < 200$ GeV/nucleon pair, is in good agreement with the one measured in $e^+e^-$ collisions vs. $\sqrt{s}$. These findings favour the hypotesis that in $AA$ collisions, at very high energy, $E_{eff}$ would coincide with $\sqrt{s_{NN}}$ for each nucleon--pair system.

Assuming that this will hold also at LHC energies, it 
is then possible, on the basis of an extrapolation 
from the $e^+e^-$ fit, to derive
a prediction for the mean total charged multiplicity per nucleon pair 
in nucleus--nucleus collisions at $\sqrt{s_{NN}}=5.5$~TeV. 
The predicted mean total charged multiplicity, 
scaled by the number of participants ($N_{part}$), would in fact be:
\begin{equation}
  \frac{2}{N_{part}}\left<n_{ch}\right>~=~72~\pm~2,
\end{equation}
resulting then in a total charged multiplicity in very central $Pb-Pb$ collisions  at the LHC (using the maximum possible value $N_{part}=416$, which is the total number of nucleons available in $Pb-Pb$ collisions):
\begin{equation}
  \left<n_{ch}\right> = 15000 \pm 400.
\end{equation}

\begin{figure}[!h]
  \begin{center}
    \small
    \includegraphics[width=0.8\linewidth]{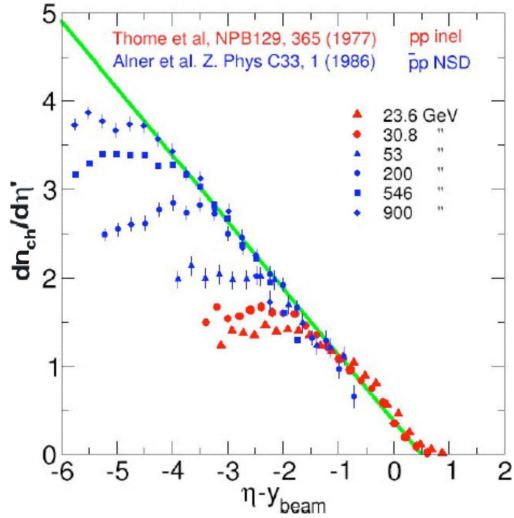}
    \caption{Limiting fragmentation in $pp$ and $p \bar{p}$ collisions at
      different centre--of--mass energies.}
    \label{ppLimitingFragmentation}
  \end{center}
\end{figure}

\begin{figure}[!h]
  \begin{center}
    \small
    \includegraphics[width=\linewidth]{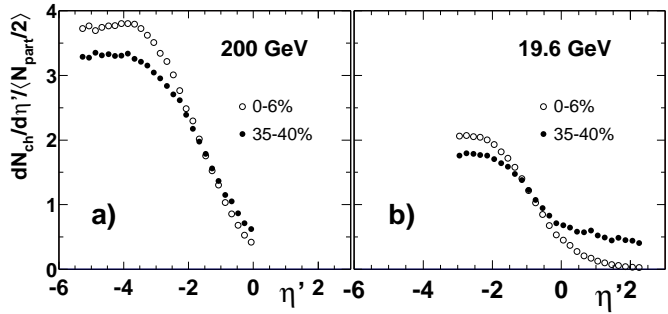}
    \caption{PHOBOS (RHIC) results for $Au-Au$ collisions at
      $\sqrt{s_{NN}} = (19.6,~200)$~GeV. In this figure, taken from \cite{AA:mult}, the results of $\frac{2}{N_{part}}dN_{ch} / d \eta'$, where $\eta' = \eta- y_{beam}$, are presented for different centrality bins.}
    \label{phobosLimitingFragmentation2}
  \end{center}
\end{figure}

Once the total multiplicity is fixed, a prediction for the charged multiplicity at midrapidity (which is relevant for the measurement of a lot of observables
in $Pb-Pb$ collisions) can be derived in the so--called  
limiting fragmentation hypothesis.
As shown in Fig.~\ref{ppLimitingFragmentation}, several measurements
 from $pp(\bar{p})$ \cite{Alner:1986xu,Thome:1977ky} at different centre--of--mass energies support the fact that the multiplicity at rapidities close 
to the beam rapidity (the so--called fragmentation region) has a shape
 which is independent of the energy in the collision.
This effect \cite{Benecke:1969sh} has been 
observed at very different centre--of--mass energies also in DIS events.
It depends only on the nature of the incident systems and, in the case 
of nuclei, on the centrality of the collision.
The same behaviour is indeed preserved in $AA$ collisions at RHIC, as shown in Fig.~\ref{phobosLimitingFragmentation2} by the measurements of the rapidity distribution for different centralities from the PHOBOS collaboration.

As shown in \cite{Alner:1986xu,Thome:1977ky,Benecke:1969sh}, the behaviour of the pseudorapidity distribution has some universal features and this allows to fix the shape
 of the pseudorapidity distribution {\it a priori}, independently of
the centre--of--mass energy of the collision.
Anyway such universal behaviour doesn't allow to establish the height at which the distribution is cut.
This information can be found requiring that the integral under the curve should give the total charged multiplicity as predicted from the extrapolation
logarithmic fit to the $e^+e^-$ data.

\begin{figure}[!h]
  \begin{center}
    \small
    \includegraphics[width=0.95\linewidth]{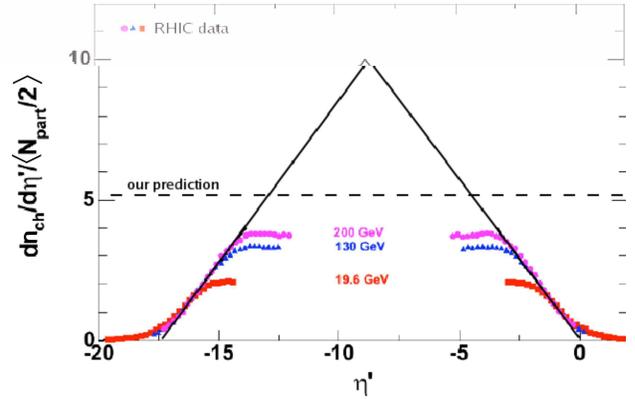}
    \caption{Limiting fragmentation at LHC based on an extrapolation from RHIC results in central (0--6\%) ion--ion collisions. The dashed line indicates the value of 
the charged multiplicity per rapidity unit (scaled by the number of participants) which results in a total multiplicity (given by the integral under the curve)
 equal to the prediction from the $e^+e^-$ fit. The rapidity of the beam at LHC is $y_{beam}~=~8.6$.}
    \label{LimitingExtrapolation}
  \end{center}
\end{figure}

Therefore, in these hypotheses it is possible to fix the shape of the
pseudorapidity distribution \cite{Alner:1986xu,Thome:1977ky,Benecke:1969sh,Back:2002wb,Back:2004je,Roland:2005su,Gelis:2006tb} and derive the mean
charged multiplicity at midrapidity in central $Pb-Pb$ collisions at LHC, as it is shown in Fig.~\ref{LimitingExtrapolation}.

The result is:
\begin{equation}
   \frac{2}{N_{part}}<\frac{dn_{ch}}{d\eta}>|_{\eta=0}~=~5.3~\pm~0.5,
\end{equation}
corresponding to about $1100~\pm~100$ charged tracks per rapidity unit.

There are a lot of predictions for this observable based on as many models.
These models, that are compatible with the actual experimental results or are tuned on them, produce a very wide spread of predictions.
In particular, a recent direct extrapolation from RHIC data seems to indicate that the charged multiplicity at LHC, in heavy ion central collisions at midrapidity, could be very small ($\left< dn_{ch}/d\eta \right> \sim 1200$) \cite{Lisa:2005js}, in good agreement with our prediction.

\begin{figure}[!h]
  \begin{center}
    \small
    \includegraphics[width=\linewidth]{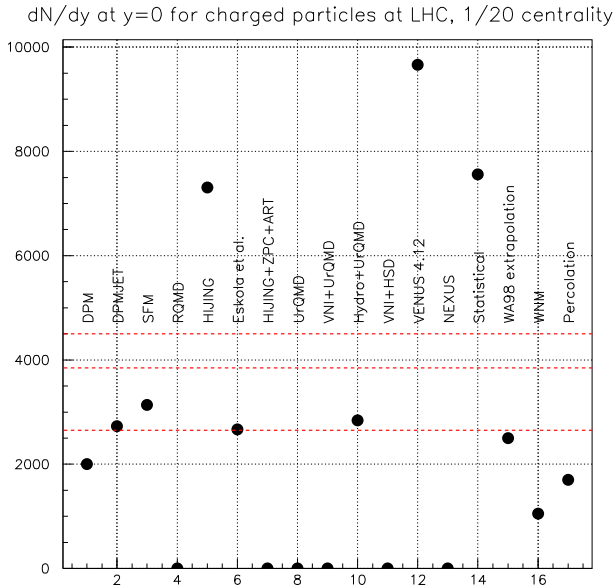}
    \caption{Mean charged multiplicity per rapidity unit at $\eta~=~0$ from different models, at $\sqrt{s_{NN}}=~5.5$~TeV. For those models for which the extrapolation to LHC energies has not been performed, the corresponding value is set to zero \cite{Armesto:2004sa}.}
    \label{MonteCarloPrevision}
  \end{center}
\end{figure}

Figure~\ref{MonteCarloPrevision} taken from \cite{Armesto:2004sa} is a review of the results of the main models.
It can be noted that there are predictions of over 6k charged tracks per rapidity unit and others of less than 3k.

Our prediction is significantly low when compared with the predictions of
widely used Monte Carlo generators (such as HIJING, for example).
It should be noted however that other effects, in particular the jet
quenching \cite{RHIC:JQ}, could sizeably increase 
the multiplicity in the central rapidity 
region.
In Fig.~\ref{MonteCarloPrevision} only the point relative to HIJING takes somehow into account such an effect (averaging on the two possible scenarios with and without quenching) \cite{Armesto:2004sa}.
An increase by a factor $\sim~2$ is actually predicted by HIJING when its default\footnote{In ALICE different variants of this algorithm are presently being investigated.} jet quenching algorithm is switched on.
Even considering this additional effect, 
if the universality features
hold at LHC energies, a relatively low value for the charged multiplicity at midrapidity is expected to be observed.
In that case, while the physics reach of ALICE will not be endangered, as illustrated in \cite{Carminati:PPR2} versus many representative physics observables, its background conditions would correspondingly improve.

\section{Conclusions}
\label{conclusion}
In this work we have estimated the average charged particle multiplicity at midrapidity of the events that will be produced in $Pb-Pb$ interactions at LHC, in an extremely high energy density domain. Our result, derived by means of detailed Monte Carlo simulations, was inspired by previous experimental findings and based on the sound hypothesis that the universality features of multihadron final states will still apply at LHC, in both $pp$ and $AA$ interactions.

The measurement of the leading baryons which will be needed to deduce the effective energy, e.g. the essential parameter of this analysis, has been proven to be feasible using the Zero Degree Calorimeters (ZDCs) of the ALICE detector.
We have shown, in particular, the capability of ALICE to perform an effective energy study in the range $3-10~{\rm TeV}$ for $pp$ collisions at $\sqrt{s} = 14~{\rm TeV}$.
This will be possible taking advantage of the acceptance and energy resolution ($\sim 500~{\rm GeV}$) of the ZDC calorimeters to measure protons and neutrons with Feynman--x $0.3 < x_{F} < 0.8$.

The rather low multiplicity prediction obtained herein for large centrality $Pb-Pb$ collisions at LHC, namely $\left< dn_{ch} / d\eta \right> = 1000-2000$ at $\eta \sim 0$ and $\sqrt{s_{NN}} = 5.5~{\rm TeV}$, adds to the so far uncertain mosaic of the predictions of such a very fundamental quantity for the description of $AA$ phenomena. \\

\end{document}